\documentclass[useAMS,usenatbib]{mn2e}
\usepackage{epsfig}
\usepackage{graphicx}
\usepackage{amssymb}
\usepackage{color}
\usepackage{placeins}
\usepackage{subfigure}
\usepackage[fleqn]{amsmath}
\usepackage{threeparttable}
\begin{document}
%%%%%%%%%%%%%%%%%%%%%%%%%%%%%%%%%%%%%%%%%%%%%%%%%%%%%%%%%
\title[Constraints on EBL using TeV Observations of BL Lacs]
  {Constraints on Extragalactic Background Light using TeV Observations of BL Lacertae objects}

\author[Qin et al.]
  {Longhua Qin$^{1}$,
    Jiancheng Wang$^{2,3}$, Quangui Gao$^{1}\thanks{qggao@yxnu.edu.cn}$, Weiwei Na$^{1}$,  \\
     \newauthor Huaizhen Li$^{1}$,Ao Wang$^{1}$, Chuyuan Yang$^{2,3}$, and Jianping Yang$^{4}$\\
     $^1$Department of Physics, Yuxi Normal University, Yuxi, Yunnan, 653100, China\\
     $^2$Yunnan Observatory, Chinese Academy of Sciences, Kunming,
Yunnan, 650011, China. \\
     $^3$Key Laboratory for the Structure and Evolution of Celestial Objects, Chinese Academy of Sciences,  Kunming, Yunnan, 650011, China.\\
     $^4$Yunnan Agricultural University, Kunming, Yunnan, 650201, China}

\pagerange{\pageref{firstpage}--\pageref{lastpage}} \pubyear{2023}

\maketitle

\label{firstpage}

%%%%%%%%%%%%%%%%%%%%%%%%%%%%%%%%%%%%%%%%%%%%%%%%%%%%%%%%%
\begin{abstract}

The extragalactic background light (EBL) in the IR to UV bands partly absorbs very high energy (VHE, $E \geq$ 100GeV) $\gamma-$ray photons travelling over cosmological distances via pair production.  In this paper, to get stronger constraints on EBL, we use the deliberate selection of the EBL  model and data of five BL Lacs with better statistics and the harder spectra to limit the EBL density and the radiation mechanism of BL Lacs. We constrain the upper limit of the EBL density by fitting the spectral energy distributions (SEDs) of TeV BL Lacs and find that our results are compatible with the published measurement, reaching 50 $\rm{nW m^{-2} sr^{-1}}$. We also obtain that the EBL is not necessarily transparent to high VHE photons. We fix the intrinsic spectral index $\Gamma_i$ of TeV BL Lacs as 1.0 and 1.5 under observation evidence and model assumption. Comparing the EBL density given by galaxy count and $Spitzer$ observations, we then obtain that 1ES 1101-232 has $\Gamma_i$ $\leq$ 1.0  and 1ES 0229+200 should have $\Gamma_i$ not harder than 1.0. We demonstrate that the common radiation assumption of BL Lacs, in which the $\Gamma_i$ is softer than 1.5, should be revisited. Furthermore, we propose that the upper EBL density could be given by fitting the hardest energy spectra of TeV BL Lacs.

\end{abstract}


%--------------------------------------------------------------------------------------------------------------------------------------------------------

\section{INTRODUCTION}
The extragalactic background light (EBL) is the diffuse photon field formed mainly by thermal processes resulting from structure formation. It exists today at wavelengths from the cosmic optical background (COB) at $\lambda_{\rm EBL}=0.1-10\mu m$ to the cosmic infrared background (CIB) at $\lambda_{\rm EBL}=10-1000\mu m$. Thus, EBL's strength and spectral shape hold fundamental information about the history of star formation and cosmological evolution \citep{ino2013,sal2021}. In addition, the EBL photons can attenuate the distant VHE (E $\geq$ 100GeV) $\gamma$-rays via the $\gamma\gamma \rightarrow e^+ e^-$ process that forms an insurmountable obstacle to the high-energy cosmic astronomy \citep{part1967,peeb1993,hau1998,dwek2013}. Therefore, the knowledge of EBL intensity and spectra is necessary and relatively important but remains controversial.

Directly measuring the EBL faces several challenges, and the most difficult task is removing the pollution of the foreground photons like zodiacal light from the Milky Way and the diffuse galactic light \citep{hau1998,hau2001,mat2006}. These attempts are engrossed in the optical \citep{ber2007,lau2020} and the IR\citep{ber2007}, while the reliability of data sets needs to be in general agreement. In addition, calibration issues also exist in practice. A strict lower limit on EBL intensity can be acquired by collecting the light emitted from resolved galaxies \citep{mad2000,hau2001}. This approach, available in many bands, including UV, optical/NIR far IR, and submillimeter, favours the local EBL determination. The acquired EBL intensity in optical and near-IR are generally below direct photometry estimates. Basic descriptions of EBL and its measurement are elaborated in the context of recent studies by \cite{sal2021}.

An indirect way to constrain the EBL is to obtain the EBL optical depth as the functions of wavelength and redshift if both the intrinsic and observational $\gamma$-rays are available from distant extragalactic sources. Blazars are the most common VHE gamma-ray sources, classified into flat-spectrum quasars (FSRQ) and BL Lacs based on the width of the emission line. Besides, blazars exist at various redshifts, and their VHE spectrum is modified by EBL absorption, becoming the natural candidate for studying EBL. This approach has been tested by several authors \citep{ste1993,dwek1994,maz2007,fin2009,yang2010,dom2013,dom2019,fer2018,desai2019,qin2020,sing2020,sing2021}, and has been remarkably successful in practice. However, this method still faces the uncertainty of the intrinsic spectral shape and optical depth model in the fitting process. In fact, the VHE $\gamma$-rays of blazars can be generated through one-zone leptonic synchrotron self-Compton model (SSC) \citep{zhang2012}, hadronic process \citep {zheng2016} or electromagnetic cascades \citep{chen2015}. The intrinsic/unattenuated energy spectra of blazars are usually assumed to constrain the EBL \citep{dom2019,fer2018,desai2019}.

Unlike previous studies focusing on each distant object (e.g., \cite{acc2019}),  we only consider three main types of TeV BL Lacs.  To get lower statistics and the FIR limits due to a pile-up at energies, we use two near and well-measured BL Lacs, Mkn421 and Mkn501. Besides, we take one distant source, 1ES0229-200, and two sources with the hard spectra, 1ES1101-232 and 1ES1218+304, because their TeV emissions are more subject to EBL absorption, and their hard spectra are natural candidates to study the lower limit on EBL. We obtain the VHE data compiled by \citep{bit2015} and derive the updated data in an online database \footnote{https://gamma-cat.readthedocs.io/}. Note that we exclude the $Fermi$-LAT data in this work since the origin of photons in the GeV bands may not be the same as in the TeV band, and fitting intrinsic spectra also requires the (quasi)-simultaneous observations.

For the BL Lacs, the intrinsic spectra in the TeV regime are characterized by $dN/dE \propto E^{-\Gamma_i}$. One can expect the same photon index  $\Gamma_i=1.5$ as the synchrotron emission spectrum if the inverse-Compton(IC) occurs in the Thomson regime and  $\Gamma_i\geq 1.5$ if the scattering happens in Klein-Nishima regime. Furthermore, if the VHE comes from the hadronic process, then $\Gamma_i=2.0$ \citep{maz2007}. However, \cite{kata2006} argue that if synchrotron emission and IC do not occur in the region of electron acceleration, the electron spectrum may become truncated due to propagation, such as $\Gamma_i<1.0$. Besides, at least some distant TeV BL Lacs, e.g., 1ES 1101-232 and 1ES 0229+200, exhibit a much harder spectral index. \cite{kren2008} point out that their spectral index $\Gamma_i$ is 1.28 or harder even under the weakest EBL model. The high density of EBL photons will lead to more absorption, requiring the harder intrinsic spectrum to get the measured spectrum that is not very soft.  Considering the above mentioned, we can set $\Gamma_i $ as 1.5 and 1.0 in the fitting procedure to determine the EBL density's lower limits.

In this paper, a flat $\Lambda$CDM cosmology with $\Omega_{\Lambda} = 0.7$, $\Omega_M = 0.3$, $H_0 = 70\rm {km\ s^{-1}\ Mpc^{-1}}$ is assumed. Besides, the unprimed and primed quantities represent the quantities in the observer and the co-moving frames respectively.

\section{MODEL AND STRATEGY }

The distant $\gamma$-ray photon of energy $E_{\gamma}$ at $z$ will interact with a background photon of energy $\epsilon$  by the cross section $\sigma_{\gamma \gamma}\left(E_{\gamma}, \epsilon, \mu\right)$:

%\begin{eqnarray}
\begin{align}
\sigma_{\gamma \gamma}\left(E_{\gamma}, \epsilon, \mu\right)=& \frac{3 \sigma_{\mathrm{T}}}{16}\left(1-\beta^{2}\right)\nonumber \\
& \times\left[2 \beta\left(\beta^{2}-2\right)+\left(3-\beta^{4}\right) \ln \left(\frac{1+\beta}{1-\beta}\right)\right],\nonumber \\
\beta \equiv & \sqrt{1-\frac{\epsilon_{\mathrm{th}}}{\epsilon}},\nonumber\\
\epsilon_{th}(E_{\gamma},\ \mu) = & {2(m_ec^2)}^2\over E_{\gamma} (1-\mu), \\
\nonumber
\end{align}
%\end{eqnarray}
where $\sigma_{\rm T}$ is the Thompson cross section, $\epsilon_{th}(E_{\gamma},\ \mu)$ is the threshold energy dictated by kinematics, $\mu$ are the cosine of the angle for the incident photons and the mass of electron, respectively. The corresponding optical depth is given by

\begin{align}\label{tau}
\tau_{\gamma}\left(E_{\gamma}, z\right)=& \int_{0}^{z}\left(\frac{d l}{d z^{\prime}}\right) d z^{\prime}\int_{-1}^{+1} d \mu \frac{1-\mu}{2} \nonumber \\
& \times\int_{\epsilon_{\mathrm{th}}^{\prime}}^{\infty} d \epsilon^{\prime} n_{\rm{EBL}}\left(\epsilon^{\prime}, z^{\prime}\right) \sigma_{\gamma \gamma}\left(E_{\gamma}^{\prime}, \epsilon^{\prime}, \mu\right),
\end{align}
where $dl/d z^{\prime}$ is the differential distant of redshift written as
\begin{equation}
\frac{dl}{dt}=c(H_0(1+z)\sqrt{(\Omega_M(1+z)^3+\Omega_{\Lambda})})^{-1},
\end{equation}
$\epsilon_{\mathrm{th}}^{\prime}=\epsilon_{\mathrm{th}}(E_\mathrm{\gamma}^{\prime},\mu)$, $E_\mathrm{\gamma}^{\prime}=E_\mathrm{\gamma}(1+z^{\prime})$, and $n_{\rm{EBL}}$ is the EBL density in the observer frame given by
\begin{eqnarray}
\epsilon^2 n_{\rm{EBL}}(\epsilon,\ z) &  = & \left( {4 \pi \over c}\right) \nu
I_{\nu}(\nu,\ z) \nonumber\\
 &  = & \int_z^{\infty}  \nu' {L}_{\nu}(\nu',\ z') \left| {{\rm d}t \over
{\rm d}z'} \right| { {\rm d}z' \over 1+z'}.
\end{eqnarray}

Following \cite{dwek2005}, we set the EBL intensity in $\rm{nW m^{-2} sr^{-1}}$ as the 9 degrees of polynomial:
\begin{equation}\label{ivh}
\log \left[\nu I_{\nu}(\lambda)\right]=\sum_{j=0}^{8} a_{j}[\log (\lambda)]^{j}.
\end{equation}

The form of the intrinsic spectrum $(\frac{dN}{dE})_{\rm int}$ used in this paper is taken from very general considerations about the emission mechanisms. The SEDs of blazars are commonly thought to be originated from the leptonic model, e.g., the SSC \citep{band1985}. In addition,  the VHE spectrum may result from hadronic scenarios like emission from leptonic models demonstrating a smooth and concave spectrum. Furthermore, the spectrum may have an addition of cut-off components caused by the Klein-Nishina effect or a cut-off in the underlying electron distribution. Considering the above reasons, the intrinsic spectrum, following \cite{hess2013},  can be expressed as a power law (PL), $N_0(E/E_0)^{-\Gamma}$, a power law with the exponential cut-off (PLE), $N_0(E/E_0)^{-\Gamma}{\rm exp}(-E/E_{cut})$, a log parabola (LP), $N_0(E/E_0)^{{\rm -a-bln}(E/E_0)}$, and a log parabola with the exponential cut-off (LPE), $N_0(E/E_0)^{{\rm -a-bln}(E/E_0)}{\rm exp}(-E/E_{cut})$. In the above formulas, $\Gamma$ and $N_0$ are the photon index and flux normalization, $a$ is the photon index at $E_0$, and $b\geq 0$ is the curvature parameter. Here, the reference energy $E_0$ is set to the decorrelation energy of the spectrum. In this work, following \cite{bit2015},  $E_0$ is fixed to the central value of the energy range of each spectrum and obtained by $\sqrt{E_{\rm {min, VHE}}E_{\rm {max, VHE}}}$.

The distant VHE $\gamma$-rays arriving at Earth can be attenuated by the EBL photons, and the observed spectrum is given by
\begin{equation}
(\frac{dN}{dE})_{\rm obs}=(\frac{dN}{dE})_{\rm int}e^{- \alpha\rm{\tau}{(E_{\rm{\gamma}},z)}},
\end{equation}
where $\alpha$ is the scale factor determined by spectral fitting.

In this paper, four widely accepted EBL templates of \cite{fra2008}(Fr), \cite{fin2009}(Fi), \cite{dom2011}(Do) and \cite{gil2012}(Gi) are used to fit each source. The first template by \cite{fra2008} is based on data sets available from $Spitzer$ IR observations and ground-based telescopes under an empirical backward evolution model to estimate EBL contribution. The distant observables at near IR bands are obtained from the contribution of the evolution of spheroidal, spiral, and merger galaxies. The rest of the EBL bands are from extrapolation (for shorter wavelengths) or the thermal dust emission from the interstellar galaxy medium (for longer IR wavelengths). The second template proposed by \cite{fin2009} predicts the intensity of EBL directly from the stellar radiation and re-emission by dust in the stellar medium. In this template, the emission in the shorter wavelength is obtained from the main and off-main sequence stars under the assumption of the star formation rate, initial mass function, and dust extinction. At the larger wavelength, the emission mainly originates from the dust, similar to a combination of three black bodies. The third template by \cite{dom2011} is mainly based on the observations of 6000 galaxies. The EBL's evolving spectrum comes from the observed rest-frame $K$-band galaxy luminosity. The final template by \cite{gil2012} is based on the semi-analytical approach describing the evolving radiation released by galaxies and quasars starting from initial cosmological conditions.

We perform the spectral fitting with a minimum chi-square given by the LMFIT package \citep{new2016} under MCMC method \citep{lew2002,mac2003,fm2013}. LMFIT is a Least-Squares Minimization routine and provides a simple and flexible approach to explore the high-level parameters' space. We input the $\chi^2$, which comes from $\chi^2=\sum_{i=1}^{N}(\frac{f_{obs}-f_{model}}{\sigma_{i}})^2$, where $f_{obs}$,$f_{model}$ are observed and modelled spectra of TeV BL Lacs, $\sigma_{i}$ is the observed error.  We thereby derive the intrinsic spectrum of a chosen source by adopting the function to obtain the lowest value of $\chi^2(\alpha)$ when $\alpha=0$. Based on the adopted model, we then fit $\alpha$ that should satisfy $\chi^2(1) \geq \chi^2(\alpha)$. We derive the final optical depth and uncertainty by calculating the mean of four scaled individual optical depths. Upon those procedures, we employ the MCMC algorithm \citep{lew2002,mac2003} to do the EBL density fitting based on Equations from Eq.\ref{tau} to Eq.\ref{ivh}.

\section{APPLICATION AND RESULTS}

\begin{figure*}
\centering
\subfigure{\includegraphics[height=19cm,width=\textwidth]{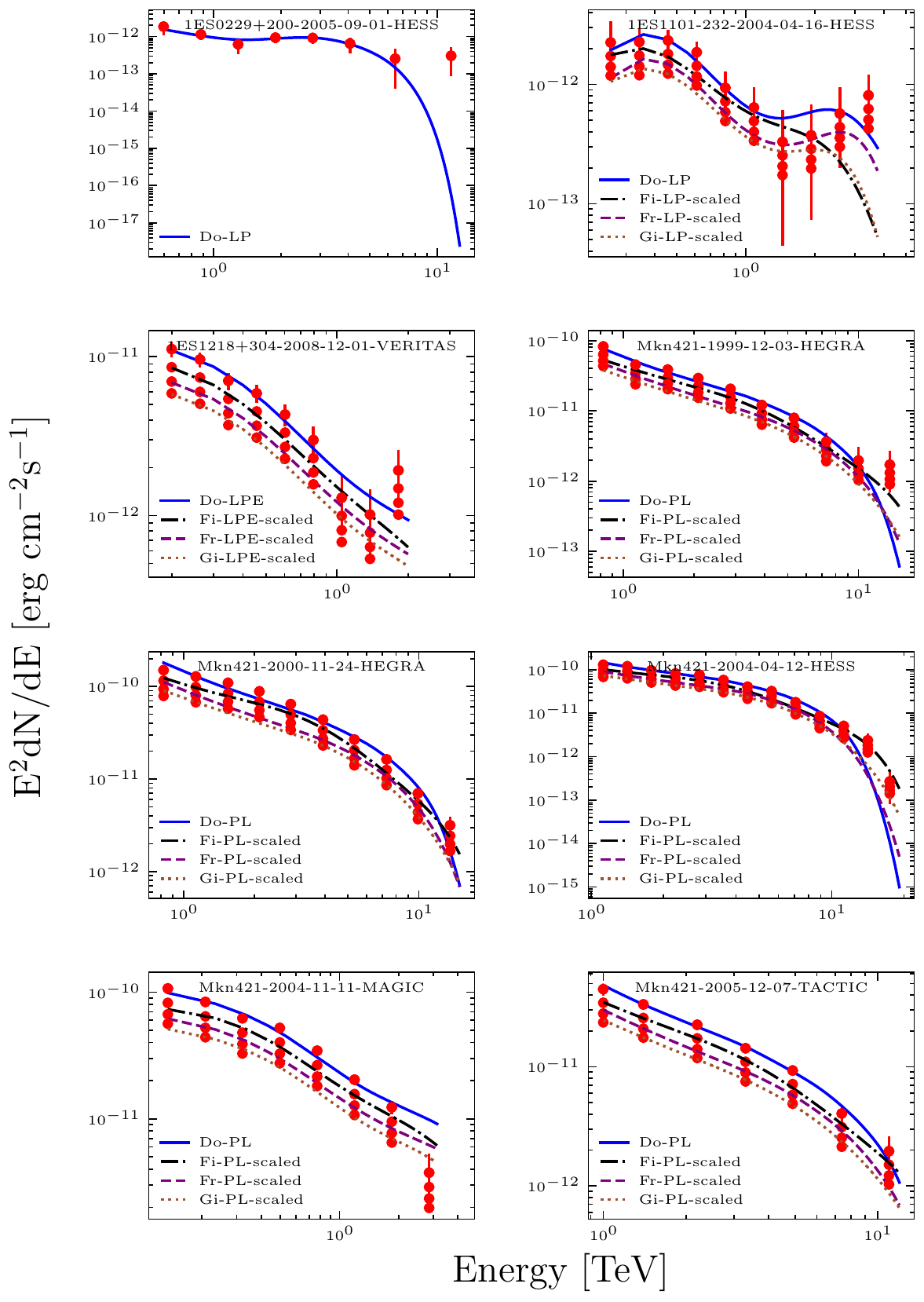}}
%\subfigure{\includegraphics[height=6.5cm,width=8.7cm]{REGECT_EBL.ps}}
\caption{Best-fit SEDs of TeV BL Lacs under four EBL templates}
 \label{Fig:1}
\end{figure*}

\begin{figure*}
\centering
\subfigure{\includegraphics[height=19cm,width=\textwidth]{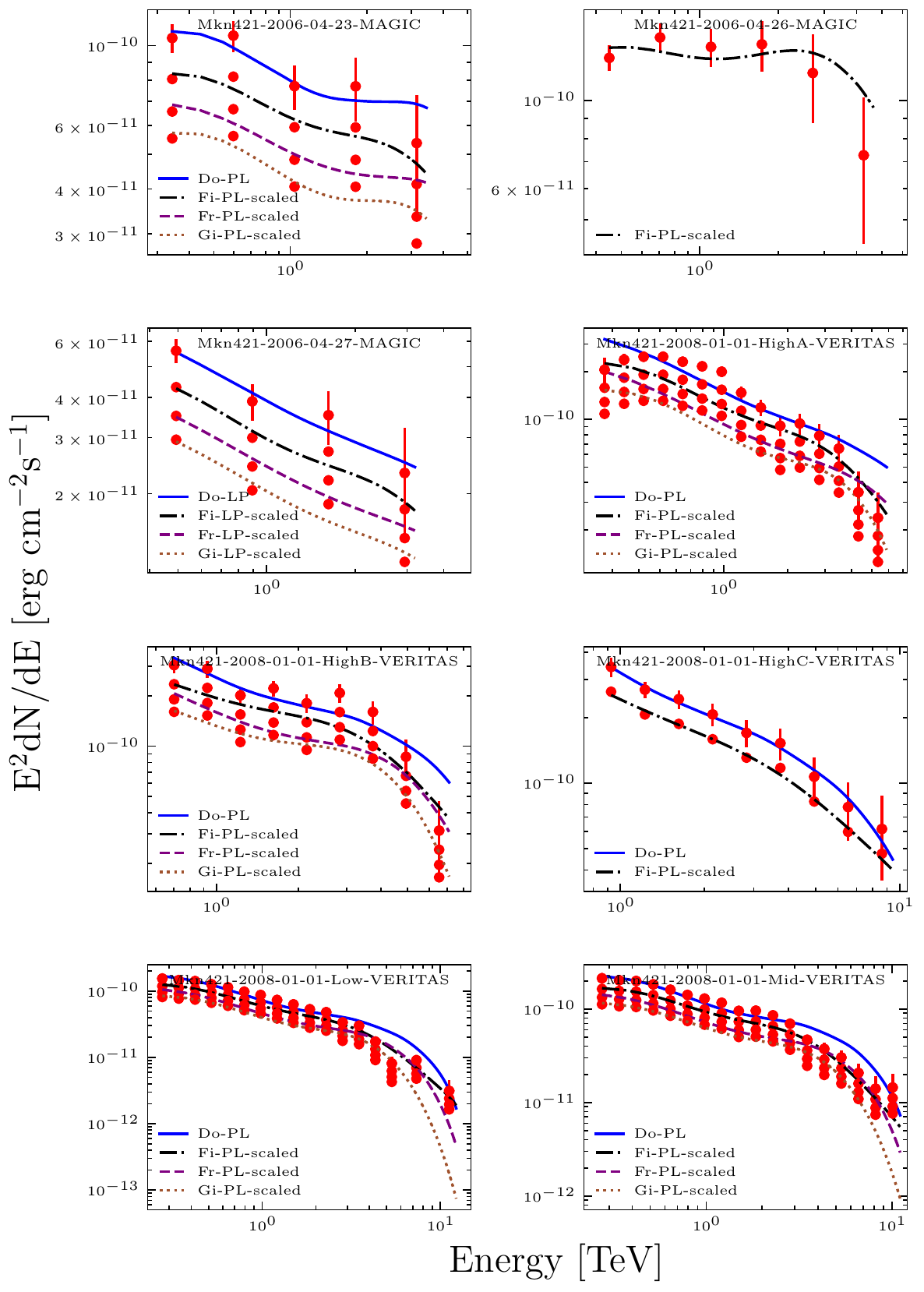}}
%\subfigure{\includegraphics[height=6.5cm,width=8.7cm]{REGECT_EBL.ps}}
\caption{Continue--As Fig.1}
 \label{Fig:2}
\end{figure*}

\begin{figure*}
\centering
\subfigure{\includegraphics[height=19cm,width=\textwidth]{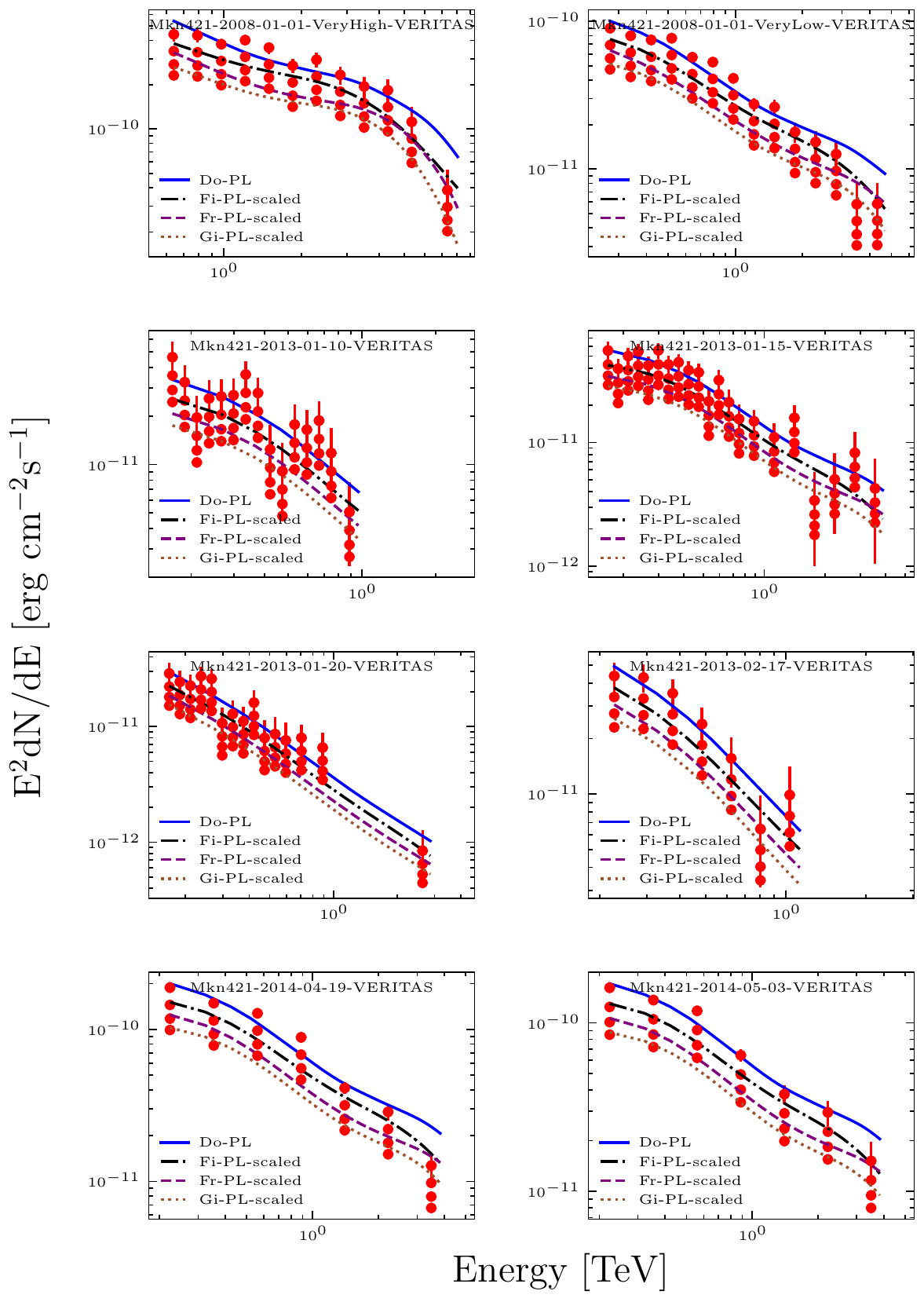}}
%\subfigure{\includegraphics[height=6.5cm,width=8.7cm]{REGECT_EBL.ps}}
\caption{Continue--As Fig.1}
 \label{Fig:3}
\end{figure*}

\begin{figure*}
\centering
\subfigure{\includegraphics[height=14cm,width=\textwidth]{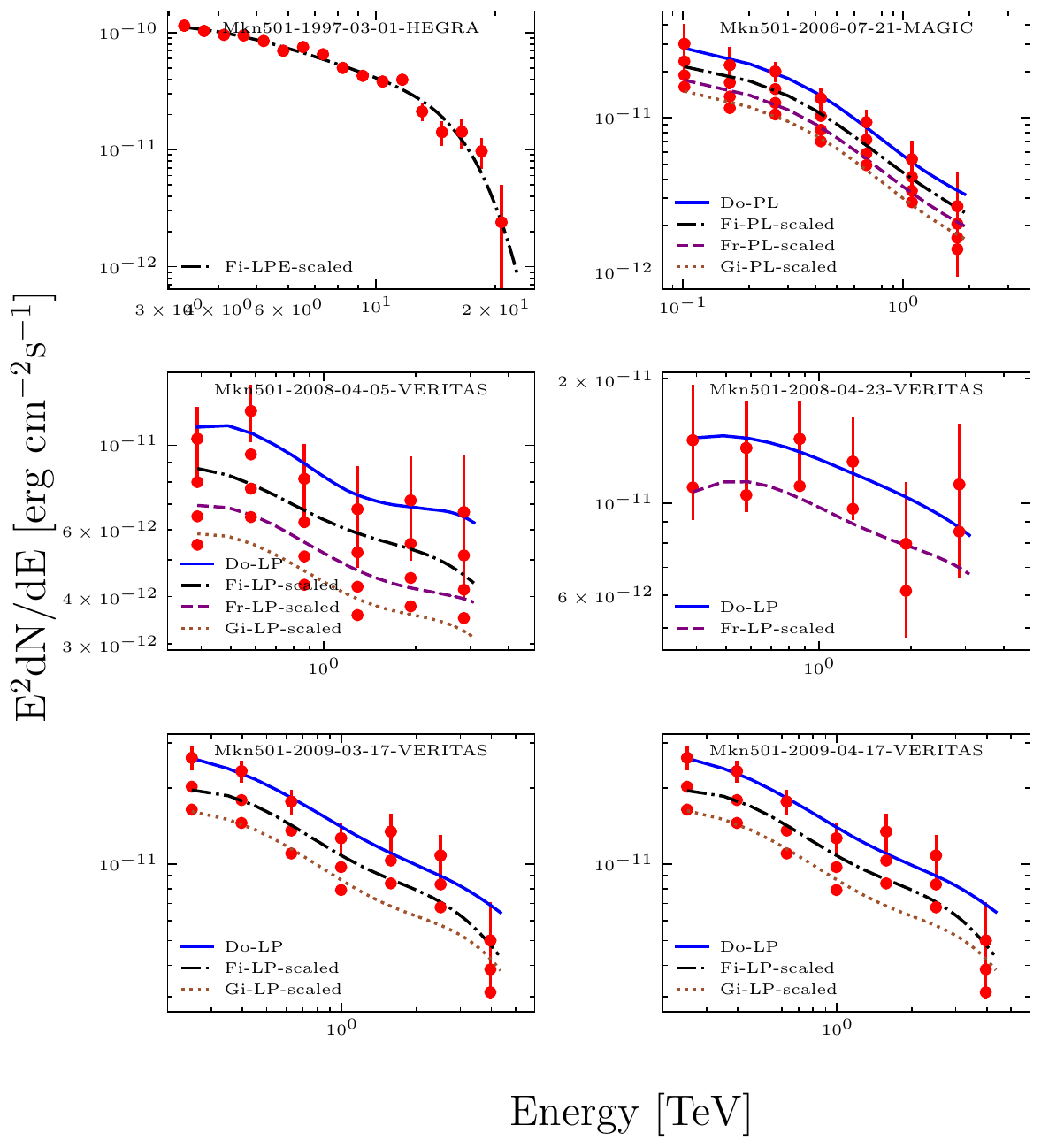}}
%\subfigure{\includegraphics[height=6.5cm,width=8.7cm]{REGECT_EBL.ps}}
\caption{Continue--As Fig.1}
 \label{Fig:4}
\end{figure*}

To get the EBL shapes at the redshift of z=0, we use the TeV data to fit the observed SEDs of 5 BL Lacs based on the assumed intrinsic spectra. At first, following the formalism described in section 2, we derive the best fitting SEDs, the scale factors, and their uncertainties. We present the results in Figure \ref{Fig:1}~-~\ref{Fig:4} and the Table \ref{Table:1}. Then we determine the nine parameters of the EBL model by fitting average scaled optical depth. We provide the best fitted spectra of the EBL and the 68\% confident level (C.L.) results in Figure \ref{Fig:5}~-~\ref{Fig:6} and the Table \ref{Table:2}. Finally, as the comparison, we also give the upper limits of the EBL density derived by direct measurements \citep{dwek2013}, $spitzer$ observations in mid-IR bands \citep{fazio2004}, galaxy counts \citep{mad2000}, and the lower limits (LW08) at 3.6$\mu m$ band \citep{leven2008}.

\subsection{Upper limits on EBL from TeV BL Lacs}

The $\gamma$-rays from the TeV BL Las attenuated by the soft photon of the EBL at redshift z via pair production process. In this paper, the VHE photons covering 100 GeV to 10 TeV emitted by TeV BL Lacs correspond to the EBL photons with 1-5$\mu m$ (0.1-8eV). The relation in dimensional units can roughly be expressed as E(TeV)$\sim$0.5$\lambda$($\mu m$), and the soft photon obtained contains only a portion of the EBL energy spectrum.

\begin{figure*}
\centering
\subfigure{\includegraphics[width=0.45\textwidth]{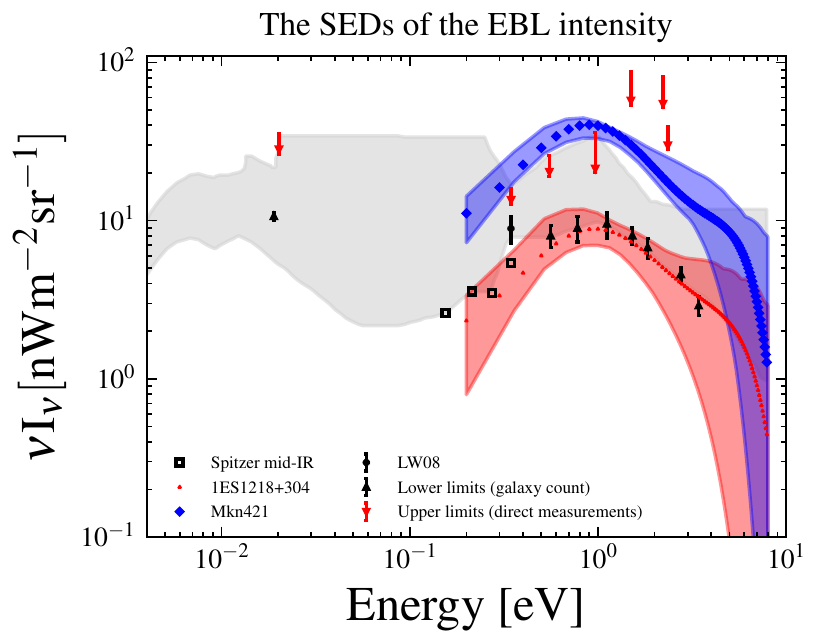}}
\subfigure{\includegraphics[width=0.45\textwidth]{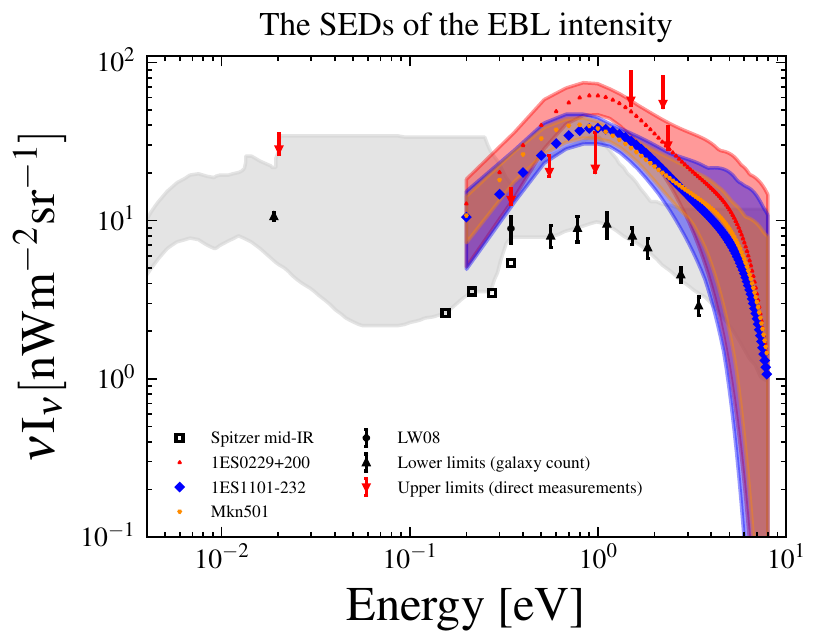}}
\caption{Spectral intensity of the EBL constrained by VHE measurements.  }
 \label{Fig:5}
\end{figure*}

The upper limits of the EBL SED $\nu I_{\nu}$ in different bands given by five TeV BL Lacs are between 10$\sim$40 $\rm{nW m^{-2} sr^{-1}}$. Figure \ref{Fig:5} shows the local EBL SED and its 68\% C.L. model uncertainties (shaded area with colors). This result is consistent with the previous studies \citep{yang2010,qin2020}. However, as shown in Figure \ref{Fig:5}, for 1ES 1218+304(z=0.18) and 1ES 0229+200(z=0.14), the derived EBL SEDs are close to or lower than the integrated ones of resolved galaxies. The results agree with those obtained by \cite{abe2019}, in which the EBL SED is calculated by reproducing the SEDs of 14 VERITAS-detected blazars under the generated EBL shapes. In addition, from Table \ref{Table:1}, we find that the optical depths obtained by MKn421(z=0.03) and Mkn501(z=0.03) are systematically higher than those obtained by the common models and the observations. The results also supported by previous studies \citep{aha2006} imply that the EBL is more transparent for the high redshift VHE photons than the local ones. However, the optical depth of 1ES 1101-232 (z=0.186) is at odds with the above conclusions and has a similar value at lower redshift.

\subsection{Lower limits on EBL from the fixed spectral index of TeV BL Lacs}

\begin{figure*}
\centering
\subfigure{\includegraphics[width=0.45\textwidth]{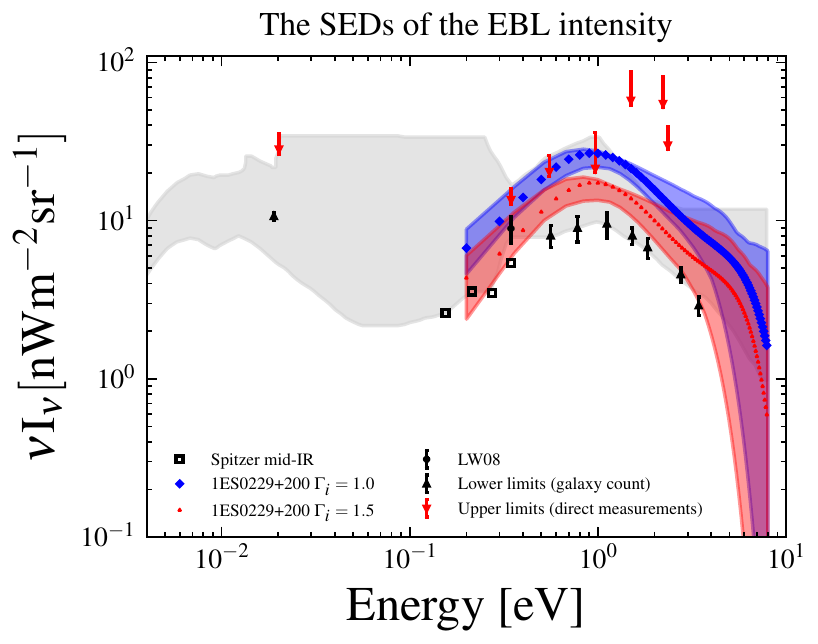}}
\subfigure{\includegraphics[width=0.45\textwidth]{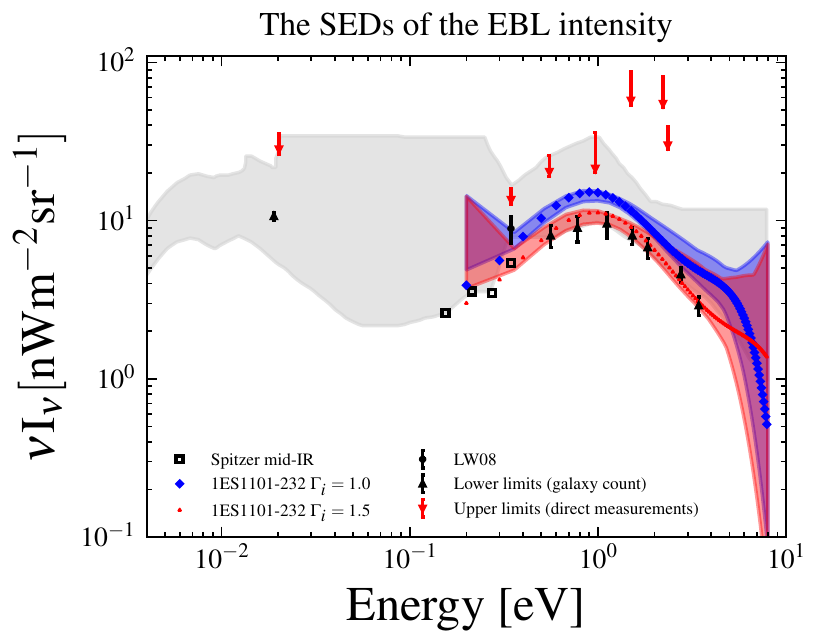}}
\caption{Spectral intensity of the EBL constrained by VHE measurements under the fixed $\Gamma$.  }
 \label{Fig:6}
\end{figure*}

The intrinsic spectra of BL Lacs follow the power-law form ($E^{-\Gamma_i}$) in the TeV band based on the theoretical analysis \citep{aha2006}. Besides, as discussed in the previous section, ${\Gamma_i}$ looks much harder than 1.5, and several spectral indices in the flaring statement are around 1.5 with large uncertainties recorded by the second $Fermi$-LAT flare catalogue \citep{abd2017}, which enable us to consider the hardest index in the study. Here, the fixed ${\Gamma_i}$ set as 1 and 1.5 have been used to detect the lower limits of the EBL intensity and test the common radiation mechanisms of the BL Lacs. The result are shown in Table \ref{Table:1}-\ref{Table:2} and Figure \ref{Fig:6}.

The harder index could require the lower EBL intensity \cite{aha2007,kren2008}, meaning that the hardest index of ${\Gamma_i}=1$ could result in the lowest limits of EBL spectra. In this paper, we have set the spectral indices of two BL Lacs as 1 and 1.5 to limit the EBL. From Table \ref{Table:1}-\ref{Table:2} and Figure \ref{Fig:6}, we can find that, for the 1ES 0229+200, the EBL SED is slightly larger than that obtained by the $Spitzer$ mid-IR and galaxy count when ${\Gamma_i}=1.5$ is fixed, while the upper limit of the derived EBL SED is close to the direct measurements if ${\Gamma_i}=1.0$ is given. Interestingly, we find that, from the left panel of Figure \ref{Fig:6}, the lower limit given by $Spitzer$ at 3.5$\mu m$ is between the upper (${\Gamma_i}=1.0$) and the lower (${\Gamma_i}=1.0$) limits of the derived EBL SED. Meanwhile, for 1ES 1101-232, we can see that the upper and lower limits of the derived EBL SED are at the top and bottom of the Finke 2010 model. However, the upper limit of the derived EBL SED is slightly lower than that measured by $Spitzer$ at the 3.5$\mu m$ band.

\begin{table*}
\centering
\Large
\caption{The scaled parameter $\alpha$ constrained by VHE data in four EBL models}
 \label{Table:1}
\resizebox{0.95\textwidth}{14mm}{
\begin{threeparttable}
\begin{tabular}{lcccc ccccc}
\hline \hline
EBL Model &Mkn 421& Mkn 501& 1ES 1218+304 & 1ES 0229+200& 1ES 0229+200(1.0)& 1ES 0229+200(1.5)& 1ES 1101-232 &1ES 1101-232(1.0)&1ES 1101-232(1.5)\\
~  & [1] & [2]  & [3] &[4]&[5]&[6] &[7]&[8]&[9] \\
\hline\hline
 Fr& 3.69$\pm$ 1.03& 4.42$\pm$ 1.64& 0.99$\pm$ 0.53& -~-& 1.48$\pm$ 0.30& 0.95$\pm$ 0.28& 4.95$\pm$ 1.61& 1.55$\pm$ 0.17& 1.15$\pm$ 0.17\\
\hline
Fi&4.50$\pm$ 0.90& 4.19$\pm$ 1.43& 0.88$\pm$ 0.55& -~-& 1.11$\pm$ 0.25& 0.72$\pm$ 0.21& 3.20$\pm$ 1.66& 1.43$\pm$ 0.16& 1.06$\pm$ 0.16\\
\hline
Do& 4.22$\pm$ 1.06& 2.34$\pm$ 1.59& 0.96$\pm$ 0.49& 2.71$\pm$ 1.62& 1.41$\pm$ 0.26& 0.92$\pm$ 0.25& 3.73$\pm$ 1.58& 1.42$\pm$ 0.15& 1.06$\pm$ 0.15\\
\hline
Gi&3.87$\pm$ 0.85& 3.71$\pm$ 1.62& 0.87$\pm$ 0.45& -~-& 1.16$\pm$ 0.30& 0.71$\pm$ 0.26& 3.42$\pm$ 1.61& 1.43$\pm$ 0.16& 1.07$\pm$ 0.16\\
\hline
 \end{tabular}
\begin{tablenotes}
%\footnotesize
\Large
\item.NOTES.--Fr,Fi,Gi,Do, four EBL templates are obtained from {\rm \cite{fra2008}},{\rm \cite{fin2009}},{\rm \cite{dom2011}} and {\rm \cite{gil2012}}, respectively.
\end{tablenotes}
\end{threeparttable}

}

%\vskip 0.4 true cm
\end{table*}

\begin{table*}

%\begin{left}
\caption{The polynomial parameters $a_{\rm j}$ derived from optical depths by Eq.\ref{ivh} }
 \label{Table:2}
\resizebox{0.95\textwidth}{75mm}{
\begin{tabular}{lc}
\hline
\hline
TeV BL Lacs &Polynomial parameters $a_{\rm j}$, j=0, 8 \\
\hline
\hline
 MRK421      &  \{1.56,  0.68, -2.37, -1.90,  4.62,  0.76, -4.64,  2.77, -0.53 \} \\
 (68\%~CL)   &  \{1.49~-~ 1.57, 0.39~-~ 0.78,-2.19~-~-0.95,-2.39~-~-0.61, \\
 ~&0.62~-~ 3.54,-1.18~-~ 2.34,-5.00~-~-1.58, 1.42~-~ 5.65,-2.25~-~-0.39\}  \\
 \hline
 MRK501      &   \{1.53,  0.71, -1.37, -2.35,  1.41,  2.78, -2.23,  0.48, -0.06\}  \\
 (68\%~CL)   &  \{1.49~-~ 1.62, 0.41~-~ 0.93,-2.50~-~-0.98,-2.57~-~-0.69, \\
 ~&0.00~-~ 3.44,-2.00~-~ 2.21,-5.00~-~-1.23, 1.76~-~ 7.16,-3.32~-~-0.66\}  \\
 \hline
 1ES 1218+304 &   \{0.93,  0.45, -2.10, -1.47,  3.49,  1.01, -3.47,  1.76, -0.29\} \\
 (68\%~CL)   &  \{0.82~-~ 0.99, 0.36~-~ 0.98,-2.83~-~-1.02,-3.00~-~-0.69, \\
 ~&0.00~-~ 3.54,-1.24~-~ 2.43,-5.00~-~-1.25, 1.94~-~ 7.41,-3.52~-~-0.75\}  \\
 \hline
 1ES 0229+200 &   \{1.76,  0.56, -2.44, -1.80,  4.11,  1.19, -4.32,  2.39, -0.43\} \\
 (68\%~CL)   &  \{1.63~-~ 1.82, 0.35~-~ 0.96,-3.14~-~-1.18,-3.00~-~-0.82, \\
 ~&0.65~-~ 3.65,-0.98~-~ 2.58,-4.22~-~-0.97, 1.80~-~ 7.49,-4.01~-~-0.78\}  \\
 \hline
 1ES 0229+200(1.0)  &   \{1.40,  0.54, -2.28, -1.48,  3.62,  0.80, -3.13,  1.56, -0.26\}\\
 (68\%~CL)   &  \{1.32~-~ 1.40, 0.30~-~ 0.67,-2.28~-~-1.05,-2.23~-~-0.50, \\
 ~&0.77~-~ 3.77,-1.03~-~ 2.51,-5.00~-~-1.68, 1.19~-~ 4.80,-1.74~-~-0.28\}  \\
 \hline
 1ES 0229+200(1.5)  &   \{1.21,  0.55, -2.48, -1.72,  4.66,  0.90, -4.60,  2.44, -0.38\}\\
 (68\%~CL)   &  \{1.11~-~ 1.21, 0.30~-~ 0.74,-2.25~-~-0.92,-2.40~-~-0.57, \\
 ~&0.00~-~ 3.55,-1.19~-~ 2.28,-5.00~-~-1.54, 1.49~-~ 6.12,-2.55~-~-0.44\}  \\
 \hline
 1ES 1101-232  &  \{1.56,  0.45, -2.28, -1.14,  3.59,  0.80, -3.61,  1.87, -0.29\}\\
 (68\%~CL)   &  \{1.46~-~ 1.60, 0.34~-~ 0.86,-2.44~-~-0.93,-2.53~-~-0.62, \\
 ~&0.00~-~ 3.55,-1.26~-~ 2.27,-5.00~-~-1.41, 1.67~-~ 6.68,-2.94~-~-0.56\}  \\
 \hline
 1ES 1101-232(1.0)  &  \{1.14,  0.64, -2.37, -2.17,  4.65,  1.47, -4.82,  2.33, -0.32\}\\
 (68\%~CL)   &  \{1.10~-~ 1.15, 0.37~-~ 0.62,-2.44~-~-1.40,-2.19~-~-0.60, \\
 ~&1.53~-~ 4.21,-0.72~-~ 2.88,-5.00~-~-1.85, 0.83~-~ 2.98,-0.78~-~-0.09\}  \\
 \hline
  1ES 1101-232(1.5)  &  \{1.03,  0.50, -2.53, -0.71,  3.62, -0.36, -2.08,  1.14, -0.17\}\\
 (68\%~CL)   &  \{0.97~-~ 1.03, 0.32~-~ 0.62,-2.38~-~-1.25,-2.10~-~-0.48, \\
 ~&1.11~-~ 4.06,-0.84~-~ 2.72,-4.44~-~-1.80, 0.88~-~ 3.47,-1.02~-~-0.13\}  \\

\hline
\end{tabular}}
%\end{left}

\end{table*}

\section{DISCUSSION AND CONCLUSION}

In this paper, we study the EBL property by fitting the observed spectra of five TeV BL Lacs, including close-by and well-studied sources,  distant sources, and hard spectral sources. We thoroughly test four shapes of intrinsic gamma-ray spectra to fit the observed SEDs using the commonly used EBL templates. Unlikely the stacking analysis \citep{fer2018,desai2019} that requires a large amount of data, we get the average scale factor $\alpha$ of EBL templates and its uncertainty by finding the lowest value of $\chi^2(\alpha)$ during the fitting procedures. From Table\ref{Table:1}, we can see that the scale factors are from 0.8 to 4.0, although their uncertainties are around 50\%, which are similar to those obtained by \citep{sin2014}. Finally, we get the EBL SED from Eq.\ref{tau} to Eq.\ref{ivh} based on the average scaled EBL templates. It is stressed that the evolution of the EBL with redshift must be considered because both the universe expansion and the evolution of radiation sources affect the EBL intensity. Therefore, to compare with the observed EBL SED, we convert the EBL SED at the redshift z to one at z=0.

We stress the following points: At first, we carefully select the samples that contain the well-observed gamma-ray data, the moderate redshifts, and the hard $\gamma$ spectra. Commonly, a source with many data can minimize the uncertainty of the derived EBL, and the appropriate distant source can balance EBL's extinction study and the re-emission of $\gamma$-rays during the propagation process \citep{zheng2016}. Secondly, the sources with hard gamma-ray spectra could reveal the lower limit of the EBL. Under simple assumptions, we can obtain the lowest limits of the EBL SED and test the radiation mechanism of TeV BL Lacs. Finally, the choice of $\Gamma$ as 1.0 and 1.5 is somewhat arbitrary. However, the previous studies revealed that the hardest index ($\sim 1.28$) is required to fit the gamma-ray spectra for some TeV BL Lacs \citep{kren2008} and the second $Fermi$-LAT flare catalogue (2FAV) could contain the sources with large uncertainties \citep{abd2017}.

We derive the final optical depth and uncertainty by calculating the mean of four scaled individual optical depths. Upon these procedures, we employ MCMC algorithm \citep{lew2002,mac2003} to fit the EBL SED based on Equations from Eq.\ref{tau} to Eq.\ref{ivh}. We can see that the maximum EBL SED is roughly consistent with the published measurement, around 50 $\rm{nW m^{-2} sr^{-1}}$. We also find that the "bump-like" spectral feature in our derived EBL SED around 1 eV band is similar to ones obtained by \citep{kor2020}, demonstrating that the excess features in EBL flux could have physical significance. Furthermore, inspired by theory and observation, we use the fixed spectral index to detect the lower limit of the EBL SED that is most likely compatible with direct observations and galaxy count. Interestingly, we know that two spectral windows mostly favour finding a faint extragalactic background; one is near the infrared band at 3.5 $\mu m$. We find that this faint value is also higher than the derived EBL density by 1ES 1101-232 under the case of $\Gamma_i=1$, implying this TeV BL Lacs may have much harder spectra. We summarize the results as following:

(1) Four TeV BL Lacs of MKn421, MKn501, 1ES1101-232, and 1ES0229+200 can well constrain the upper limits of EBL SED. However, for 1ES0229+200 with lower $\alpha$, the EBL SED is higher than that obtained by another source with large scale factor, such as 1ES1101-232, because the fitting optical depths have large uncertain (the error to be $\geq$ 1.6) and the EBL templates are only given by \cite{fin2009};

(2) The upper limits of EBL density is around 50 $\rm{nW m^{-2} sr^{-1}}$. at 1 eV, and the lower density could be at ones between given by $spitzer$ \citep{fazio2004} and galaxy counts \citep{mad2000};

(3) 1ES1218+304 favours the scenario that the EBL is more transparent to VHE $\gamma$-rays. However, 1ES1101-232 with a similar redshift does not favour this collusion;

(4) Using the fixed spectral index to limit the EBL, for the 1ES1101-232, we doubt that its spectral index could be harder than that we considered before and below 1.0.

From above, the behavior of EBL is still complex, and the EBL SEDs constrained by TeV objects vary with redshifts and directions from our results. \cite{mir2009} singled out that axion-like particles (ALPs) could affect the EBL distribution in different sky directions. However, ALPs work at the redshift z$>$0.2 \citep{mir2009}. The EBL inhomogeneity could exist at only the order of 1\%, which is much smaller than the results in this paper. The proper reason is that the evolution of the EBL with redshift must be fully considered. In fact, the evolution of the EBL number density  $n_{\rm EBL}$ is  scaled redshift as $(1+z)^{3-k}$ rather than the empirical $(1+z)^{3}$, and \cite{sing2021} point out that $k$ relates to redshift. Another mechanism to explain our results is that the origin of TeV $\gamma$-rays is still unclear, the intrinsic spectrum used here does not fully cover, and the VHE could be much harder than that in previous studies when the spectral index is fixed at 1.0. Besides, the recent VHE observations of distant gamma-ray sources show that the EBL is more transparent to $\gamma$-rays than we thought before.

%______________________________________________________________
\section*{Acknowledgments}
The authors would like to thank the anonymous referee for the helpful comments,  which improved the manuscript significantly.   The authors gratefully acknowledge the financial supports from the National Natural Science Foundation of China (grants 12063005, 12063006 and U2031111), the Special Basic Cooperative Research Programs of Yunnan Provincial Undergraduate Universities (grants 2019FH001-012, 2019FH001-076 and 202001BA070001-031), the Program for Innovative Research Team (in Science and Technology) in University of Yunnan Province (IRTSTYN), the Science Research Foundation of Yunnan Education Department of China (grants 2020J0649), the program for Reserve Talents of Young and Middle-aged Academic and Technical Leaders in Yunnan Province (grants 202205AC160087) and the National Natural Science Foundation of Yunnan Province (grants 202101AU070010). The authors (QLH) gratefully acknowledge the financial supports from the Hundred Talents Program of Yuxi (grants 2019).

\section*{Data availability}
The data underlying this article are derived from gammacat: https://gamma-cat.readthedocs.io, and will be shared on reasonable request to the corresponding author.

%%\clearpage

%++++++++++++++++++++++++++++++++++++++++++++++++++++++++++++++++++++++++++++++++

%\end{document}
\clearpage
\end{document}